\documentclass[preprint,aps,superscript,address,showkeys,showpacs,pre]{revtex4}
\usepackage{amsmath,amssymb}
\usepackage{epsfig}
\usepackage{graphics}
\usepackage{graphicx}
\usepackage{float}
\usepackage{dcolumn}

\begin{document}
\title{Vibrational Resonance in the Duffing Oscillator with Distributed Time-Delayed Feedback}

\author{C.~Jeevarathinam}
\affiliation{School of Physics, Bharathidasan University, 
Tiruchirappalli  620 024, India}

\author{S.~Rajasekar}
\email{rajasekar@cnld.bdu.ac.in}
\affiliation{School of Physics, Bharathidasan University, 
Tiruchirappalli  620 024, India}

\author{M.A.F.~Sanju\'{a}n}
\email{miguel.sanjuan@urjc.es}
\affiliation{Nonlinear Dynamics, Chaos and Complex Systems Group, Departamento de F\'{i}sica \\ Universidad Rey Juan Carlos, Tulip\'{a}n s/n, 28933 M\'{o}stoles, Madrid, Spain}

\begin{abstract}
We analyze the vibrational resonance in the Duffing oscillator system in the presence of (i) a gamma distributed time-delayed feedback and (ii) integrative time-delayed (uniformly distributed time delays over a finite interval) feedback. Particularly, applying a theoretical procedure we obtain an expression for the response amplitude $Q$ at the low-frequency of the driving biharmonic force. For both double-well potential and single-well potential cases we are able to identify the regions in parameter space where either (i) two resonances, (ii) a single resonance or (iii) no resonance occur. Theoretically predicted values of $Q$ and the values of a control parameter at which resonance occurs are in good agreement with our numerical simulation. The analysis shows a strong influence of both types of time-delayed feedback on vibrational resonance.
\vskip 20pt

Accepted for publication in J. Applied Nonlinear Dynamics (in 2015)
 
\end{abstract}
\maketitle
\newpage

\section{Introduction}
Time-delay is thought to be ubiquitous in many realistic models in physics, engineering and biology. When the state of a system at time $t$ depends on its state values at previous past time, then a time-delayed feedback term is introduced in the evolution equation of the system. We can think about two main types of time-delays: discrete time-delays and distributed time-delays {\cite{ref1}}. In the present work we are concerned with distributed time-delays.

The general form of a distributed delay feedback term (DFT) in a dynamical system is given by 
\begin{equation}
\label{eq1}
  {\mathrm{DFT}} = \int_0^{\infty} G(\tau)  
         x(t-\tau)\, {\mathrm{d}} \tau,
\end{equation}
where $x$ is a state variable of the system and $G(\tau)$ is a distributed delay kernel with $G(\tau) \ge 0$ and $\int_0^{\infty} G(\tau) \, {\mathrm{d}} \tau=1$. The choice $G$ as the Dirac-delta distribution $\delta(\tau-\alpha)$ gives
\begin{equation}
\label{eq2}
   \mathrm{DFT} = \int_0^{\infty} \delta(\tau-\alpha)  x(t-\tau) 
        \, {\mathrm{d}} \tau = x(t-\alpha),
\end{equation}
that is, the delay time is a constant. The $G(\tau)$ being
\begin{equation}
\label{eq3}
    G_{\mathrm{u}}(\tau)
       = \left\{ \begin{array}{ll}
         {1/\alpha},  & \quad \tau \in [0,\alpha] \\
              0, & \quad {\mathrm {otherwise}}  \end{array} \right.
\end{equation}
leads to  ${\mathrm{DFT}} = (1/\alpha) \int_0^{\alpha} x(t-\tau)\, {\mathrm{d}} \tau$ which under a change of variable $t-\tau=t'$ becomes 
\begin{equation}
\label{eq4}
   \mathrm{DFT} = \frac{1}{\alpha} \int_{t-\alpha}^{t} x(t') 
                    \, {\mathrm{d}} t'.
\end{equation}
The kernel in this case is uniformly distributed and the delay feedback term given by Eq.~(\ref{eq4}) is known as an integrative time-delay term {\cite{ref2,ref3,ref4,ref5}}. The parameter $\alpha$ represents the width and amplitude of the uniform distribution $G_{\mathrm{u}}$. The mean and variance of the time-delay in the distribution $G_{\mathrm{u}}$ is ${\alpha/2}$ and ${\alpha^2/12}$, respectively. $\alpha$ can be treated as the response time of a system. The integrative time delay was earlier considered in the `integrate-and-fire' models {\cite{ref2}} and self-organized criticality {\cite{ref3}}. This time-delay gives rise to amplitude death {\cite{ref4,ref6,ref7}} in coupled oscillators and multiple resonance curves with and without hysteresis {\cite{ref5}}.

Another form of distributed delays which is shown to have a strong influence on the dynamics of a system is the gamma distribution of delays given by 
\begin{equation}
\label{eq5}
     G_{\mathrm{g}}(\tau)
          = \frac{\tau^{p-1} \alpha^p {\mathrm{e}}^{-\alpha \tau } }
               {\Gamma(p)} \, ,   
\end{equation}
where $\alpha$, $p \ge 0$ and $\Gamma(p)$ is the Euler gamma function. For $p=1$ the distribution $G_{\mathrm{g}}$ becomes an exponential distribution. For the gamma distribution $\langle \tau \rangle ={p/\alpha}$ while $\sigma^2={p/\alpha^2}$.

The effect of distributed delays has been reported in a multitude of scientific problems. Among them, we refer on stability {\cite{ref8,ref9,ref10}} and emergence of chaos {\cite{ref11}} in neural networks, Hopf bifurcation and stabilization of a fixed point in a discrete logistic model {\cite{ref12}}, amplitude death in coupled oscillators {\cite{ref13,ref14}}, global asymptotic behavior of a chemostat model system {\cite{ref15}}, existence of wavefront solutions in reaction diffusion systems {\cite{ref16,ref17}}, bifurcation analysis in an epidemic model {\cite{ref18}}, transmission dynamics of malaria {\cite{ref19}}, dynamics of car-following model {\cite{ref20}}, first passage time statistics {\cite{ref21}} and stochastic process in a Langevin equation {\cite{ref22}}.

Studies of different nonlinear phenomena in time-delayed systems have received a great interest in recent years. And the main goal of this paper is to investigate the role of gamma distributed time-delayed and integrative time-delayed feedbacks on the vibrational resonance produced in a Duffing oscillator, taken as the reference model system. Vibrational resonance is a resonant dynamics induced at the low-frequency $\omega$ of the input periodic signal by a relatively high-frequency of the input signal {\cite{ref23,ref24,ref25}}. The effect of a single constant and multiple constant time-delays on vibrational resonance has been previously reported {\cite{ref26,ref27,ref28,ref29}}. When the delay time is not constant or not priorly known, then it is more realistic to consider distributed delays, and this is what we do here.

The paper is organized as follows. In sec.~2 we consider the Duffing oscillator driven by a biharmonic force with two frequencies $\omega$ and $\Omega$ with $\Omega \gg \omega$ and with a gamma distributed time-delayed feedback term. We denote $\gamma$ as the strength of the feedback term. A gamma distributed time-delayed feedback is characterized by parameters $\gamma$, $\alpha$ and $p$ (refer Eq.~(\ref{eq5})). Assuming that the actual motion of the system consists of a slow motion and a fast motion and applying a theoretical method, we construct an analytical expression for the response amplitude $Q$ (ratio of the amplitude of the slow motion of the output and the amplitude of the low-frequency component of the input biharmonic driving force). We analyze the double-well potential and the single-well potential cases separately. We treat the amplitude $g$ of the high-frequency force as a control parameter. From the obtained theoretical expression of $Q$, we obtain the expression for $g$ at which resonance occurs. For the double-well potential case, we identify the regions in the $(\gamma, p)$ parameter space where either two resonances or just one resonance occur. We point out the mechanism of occurrence of the resonance in terms of the resonant frequency. In the single-well case, we show that at most only one resonance can occur. We identify the conditions on $\gamma$ and $p$ for a single resonance and no resonance. We show that all the theoretical results are in good agreement with our numerical simulations. In sec.~3 we analyse the influence of the integrative time-delayed feedback term in the Duffing oscillator. The integrative time-delayed feedback term is characterized by the time-delay parameter $\alpha$ (Eq.~(\ref{eq4})) and the strength $\gamma$ of the feedback. For this feedback type, we are also able to obtain a theoretical expression for $Q$. We report the effect of the parameters $\gamma$ and $\alpha$ on vibrational resonance in detail for both, the double-well potential and the single-well potential system separately. The final section contains concluding remarks.

\section{Duffing Oscillator with a Gamma Distributed Time-Delayed Feedback}         
The dynamics of the Duffing oscillator subjected to a biharmonic force and a gamma distributed time-delayed feedback term is governed by the equation of motion
\begin{equation}
\label{eq6}
  \ddot{x} + d \dot{x} + \omega_0^2 x + \beta x^3 
      + F(\tau,x(t-\tau))  = f \cos \omega t + g \cos \Omega t,
\end{equation}
where $\Omega \gg \omega$ and $F$ is the gamma distributed time-delayed feedback term given by 
\begin{equation}
\label{eq7}
     F(\tau,x(t-\tau)) = \gamma \int_0^{\infty} 
          G_{{\mathrm{g}}}(\tau) \, x(t-\tau) \, {\mathrm{d}} \tau
\end{equation}
with $G_{{\mathrm{g}}}(\tau)$ given by Eq.~(\ref{eq5}). For simplicity, we fix $\alpha=1$.

\subsection{Theoretical Estimation of the Response Amplitude}
We can determine the solution of (\ref{eq6}) in the long time limit for $\Omega \gg \omega$ by writing $x(t)=X(t)+\psi(t,\Omega t)$ where $X$ and $\psi$ are the slow and the fast variables, respectively, and $\langle \psi \rangle = (1/2 \pi) \int_0^{2 \pi} \psi \, {\mathrm{d}} \tau=0$. The evolution equations for $X$ and $\psi$ are
\begin{subequations}
 \label{eq8}
\begin{eqnarray}
    & & \ddot{X} + d \dot{X} + \left( \omega_0^2 
          + 3\beta \langle \psi^2 \rangle \right) X 
          + \beta \left( X^3 + \langle \psi^3 \rangle \right) 
            \nonumber \\
    & & \quad \quad + F(\tau,X(t-\tau)) = f \cos \omega t, \\
    & & \ddot{\psi} + d \dot{\psi} + \omega_0^2 \psi 
          + 3 \beta X^2 \psi  
          + 3 \beta X \left( \psi^2 - \langle \psi^2 \rangle \right)
          \nonumber \\
    & & \quad \quad + \beta \left( \psi^3 -\langle \psi^3 \rangle \right) 
          + F(\tau,\psi(\Omega t- \Omega \tau)) 
          = g \cos \Omega t. 
\end{eqnarray}
\end{subequations}
As $\psi$ is a rapidly varying function of time, it is reasonable to approximate Eq.~(\ref{eq8}b) as 
\begin{equation}
\label{eq9}
   \ddot{\psi} + F(\tau,\psi(\Omega t - \Omega \tau))
       = g \cos \Omega t.
\end{equation}
For $t \to \infty$, we assume the solution of Eq.~(\ref{eq9}) as $ A_{\mathrm{H}} \cos( \Omega t + \phi)$. The second term in Eq.~(\ref{eq9}) is then worked out as (for details see Appendix $A$)
\begin{subequations}
\begin{eqnarray}
 \label{eq10}
      F(\tau,\psi(\Omega t- \Omega \tau))
         = A_{\mathrm{H}} \,\Omega^{-p} \, \cos(\Omega t + \theta + \phi),      
\end{eqnarray}
where
\begin{eqnarray}
    \theta = \frac{p}{\Omega} - \frac{p\pi}{2} \, .
\end{eqnarray}
\end{subequations}
Substituting $\psi = A_{\mathrm{H}} \cos( \Omega t + \phi)$ and $F$ given by Eq.~(\ref{eq10}) in Eq.~(\ref{eq9}), we obtain
\begin{subequations}
 \label{eq11}
\begin{eqnarray}
 A_{\mathrm{H}} 
     & = & \frac {g}{\mu}, \quad   \mu =  \left[ {\left(\Omega^2
             - \gamma  \Omega^{-p} \cos \theta \right)^2
             + \left( \gamma  \Omega^{-p} 
                 \sin \theta \right)^2}\right]^{1/2} , \\
    \phi & = & \tan^{-1} \left( \frac{\gamma  \Omega^{-p} \sin \theta}
             {\Omega^2 - \gamma  \Omega^{-p} \cos \theta } \right).
\end{eqnarray}
\end{subequations}  
Further, $\langle \psi \rangle = 0$, $\langle \psi^2 \rangle = (1/2 \pi) \int_0^{2 \pi} \psi^2 \, {\mathrm{d}} \tau = A^2_{\mathrm{H}} / 2$ and $\langle \psi^3 \rangle = 0$. Then the equation of motion of the slow variable $X$ given by Eq.~(\ref{eq8}a) becomes
\begin{equation}
 \label{eq12}
    \ddot{X} + d \dot{X} + C_1 X + \beta X^3 
     + F(\tau,X(t-\tau)) = f \cos \omega t, 
\end{equation}
where $C_1 = \omega_0^2 +  {3\beta g^2}/(2\mu^2)$.

Slow oscillations occur about the equilibrium points $X^*$ of Eq.~(\ref{eq12}) with $f=0$.  To determine $X^*$, we write $X(t-\tau)=X(t)=X^*$. In this case $F$ in Eq.~(\ref{eq12}) is simply $\gamma  X^*$ since $ \int_0^{\infty} \tau^{p-1} {\mathrm{e}}^{-\tau} {\mathrm{d}} \tau = \Gamma (p)$. The equilibrium points are $X^*_0 = 0$, and $X^*_\pm = \pm \sqrt{-{(C_1+\gamma)}/{\beta}}$.

For convenience, we introduce the change of variable $Y=X-X^*$ and obtain
\begin{equation}
\label{eq13}
    \ddot{Y} + d \dot{Y} + \omega_{\mathrm{r}}^2 Y
      + 3 \beta X^* Y^2  + \beta Y^3 + F(\tau,Y(t-\tau))
       = f \cos \omega t , 
\end{equation}
where $\omega_{\mathrm{r}}^2 = C_1 + 3 \beta X^{*2}$. $\omega_{\mathrm{r}}$ is termed as the resonant frequency of the slow motion. For $\vert f \vert \ll 1$ and  $\vert Y \vert \ll 1$ we can drop the nonlinear terms in Eq.~(\ref{eq13}) and write its solution in the limit $t\to \infty$ as $Y=A_{\mathrm{L}}\cos(\omega t+\Phi)$. For this solution, referring to the Eqs.~(\ref{eqA1}) and (\ref{eqA3}), the distributed delays feedback term $F$ is written as 
\begin{equation}
\label{eq14}
    F = {\mathrm{Re}} \left[ \frac{\gamma A_{\mathrm{L}} 
         {\mathrm{e}}^{{\mathrm{i}}{(\omega t + \Phi )}}}
              {(1+{\mathrm{i}} \omega)^p} \right] .
\end{equation}
In Eq.~(\ref{eqA3}) of Appendix $A$, the term $1/(1+{\mathrm{i}} \Omega)^p$ is approximated as ${\mathrm{i}}^{-p}  \Omega^{-p} \, {\mathrm{e}}^{{\mathrm{i}}p/\Omega}$ assuming that $\Omega \gg 1$. We cannot use this kind of approximation in Eq.~(\ref{eq14}) because $\omega$ can be $<1$. However, defining $a+{\mathrm{i}}b = 1/(1+{\mathrm{i}} \omega)^p$, we rewrite Eq.~(\ref{eq14}) as 
\begin{equation}
\label{eq15}
    F = \gamma A_{\mathrm{L}} \left[ a \cos(\omega t + \Phi ) 
          - b \sin (\omega t + \Phi) \right].
\end{equation}
Following the procedure used to determine $A_{\mathrm{H}}$, we obtain
\begin{equation}
\label{eq16}
    A_{\mathrm{L}} 
      = {f/\sqrt{S}}, \quad
          S = \left( \omega_{{\mathrm{r}}}^2
          - \omega^2 + \gamma a \right)^2
          + \left( d \omega + \gamma b \right)^2 .         
\end{equation}
Then, we define the response amplitude $Q=A_{\mathrm{L}}/f={1/\sqrt{S}}$.

\subsection{Resonance Analysis}
We analyse the occurrence of a resonance in both the double-well and the single-well potentials of the system. The potential of the system in absence of damping, feedback term and the external periodic force is of a double-well form for $\omega_0^2 < 0$, $\beta>0$ and of a single-well form for $\omega_0^2$, $\beta > 0$.

We numerically compute $Q$, to verify the theoretical expression $Q$. For this purpose, we integrate Eq.~(\ref{eq6}) using the Euler method with step size $0.01$. Leaving the solution corresponding to first $10^3$ drive cycles as a transient, we compute the sine and cosine components of $Q$, denoted as $Q_{\mathrm{s}}$ and $Q_{\mathrm{c}}$, respectively, using the formula
\begin{subequations}
 \label{eq17}
\begin{eqnarray}
    Q_{{\mathrm{s}}} 
       & = &  \frac{2}{nT} \int_0^{nT} x(t) \sin \omega t
               \, {\mathrm{d}} t , \\
    Q_{{\mathrm{c}}} 
        & = & \frac{2}{nT} \int_0^{nT} x(t) \cos \omega t
                \, {\mathrm{d}} t ,
\end{eqnarray}
\end{subequations}
where $n=10^3$ and $T = 2 \pi / \omega$. Then, $Q = \sqrt{ Q^2_{{\mathrm{s}}} + Q^2_{{\mathrm{c}}}} {\big{/}} f$.

\begin{figure}[t]
\begin{center}
\epsfig{figure=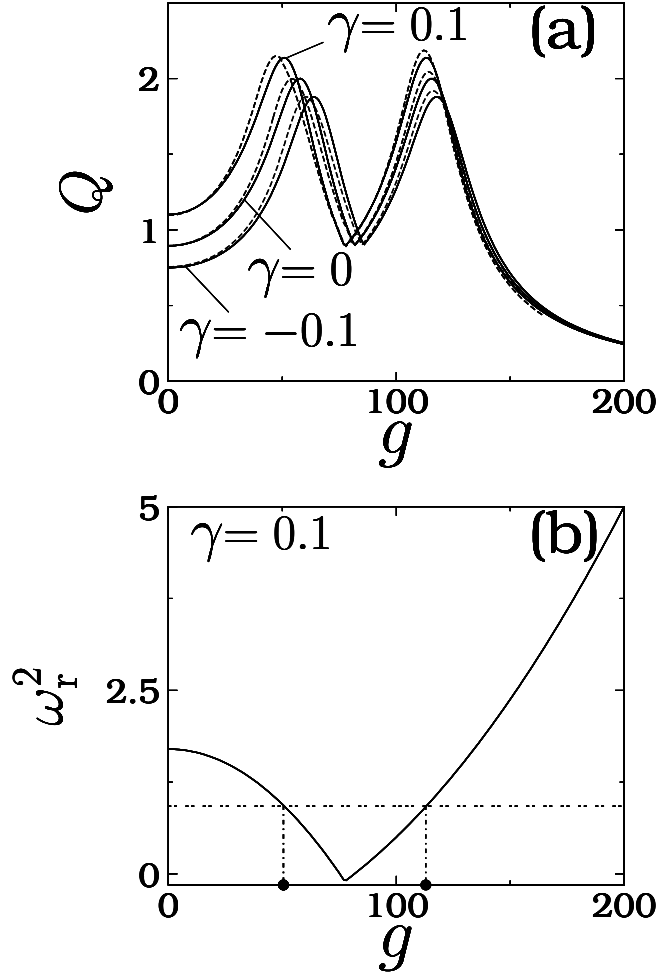, width=0.53\columnwidth}
\end{center}
\caption{(a) $Q$ versus $g$ for different values of $\gamma$ for the system {(\ref{eq6})} with a double-well potential. The continuous and dashed lines are theoretical and numerical results, respectively. (b) $\omega_{\mathrm{r}}^2$ as a function of $g$. The horizontal dashed line corresponds to  $\omega^2 - \gamma a$.  The two solid circles mark the values of $g$ at which resonance occurs. The values of the parameters of the system {(\ref{eq6})} are $d=0.5$, $\omega_0^2=-1$, $\beta=1$, $f=0.1$, $\omega=1$, $\Omega=10$, and $p=0.5$.}
 \label{f1}
\end{figure}

For the case of the double-well potential system, we fix the values of the parameters as $d=0.5$, $\omega_0^2=-1$, $\beta=1$, $f=0.1$, $\omega=1$, $\Omega=10$ and $p=0.5$. In Fig.~\ref{f1}a both theoretically and numerically calculated $Q$ are plotted as a function of the control parameter $g$ for three fixed values of $\gamma$. 
The theoretical $Q$ closely matches with the numerically computed $Q$. For a wide range of values of $g$, the response amplitude for $\gamma=0.1$ is higher than that of $Q(\gamma=0)$ , while $Q(\gamma=-0.1)<Q(\gamma=0)$. In Fig.~\ref{f1}a $Q$ becomes maximum at two values of $g$. In order to capture the mechanism of the observed vibrational resonance in Fig.~\ref{f1}b, the variation of   $\omega_{{\mathrm{r}}}^2 = C_1+3 \beta X^{*2}$ with $g$ is shown for $\gamma=0.1$. For $g<g_{\mathrm{c}}$,  where 
\begin{equation}
 \label{eq18}
    g_{\mathrm{c}}
      = \left[ \frac{2\mu^2}{3\beta}            
           \left( \vert \omega_0^2 \vert - \gamma  \right)
            \right]^{1/2}, \quad  \gamma < \vert \omega_0^2 \vert ,
\end{equation}
there are three equilibrium points  $X^*_0 = 0$ and  $X^*_\pm$. Slow oscillations take place about  $X^*_\pm$. For $g>g_{\mathrm{c}}$  there is only one  equilibrium point $X^*_0$ and a slow motion occurs about it. In the calculation of $\omega_{{\mathrm{r}}}^2$, we used $X^*=X^*_\pm=\pm \sqrt{-(C_1+\gamma)/\beta}$ for $g<g_{\mathrm{c}}$ and $X^*=X^*_0 = 0$ for $g>g_{\mathrm{c}}$.

In the expression for $A_{\mathrm{L}}$ given in Eq.~(\ref{eq16}) when $g$ is varied, the quantity $\omega_{{\mathrm{r}}}^2$ alone varies while all other terms remain the same. We notice that $Q$ or $A_{\mathrm{L}}$ becomes a maximum when the quantity $S$ becomes a minimum. This happens whenever $\omega_{{\mathrm{r}}}^2$ matches with $\omega^2-\gamma a$. In Fig.~\ref{f1}b for $\gamma=0.1$ as $g$ increases from $0$ the value of $\omega_{{\mathrm{r}}}^2$ decreases from $2 \vert \omega_0^2 \vert - 3 \gamma=1.7$ and becomes the minimum value $- \gamma$ at $g=g_{\mathrm{c}}$. When $g$ is further increased, $\omega_{{\mathrm{r}}}^2$ increases monotonically. At two values of $g$, denoted as $g_{_{\mathrm{VR}}}^{(1)}$ and  $g_{_{\mathrm{VR}}}^{(2)}$, $\omega_{{\mathrm{r}}}^2=\omega^2-\gamma a$ (marked by the horizontal dashed line in Fig.~\ref{f1}b). At these values of $g$, the response amplitude attains the maximum value ${1/(d \omega + \gamma b)}$. For $\gamma=0.1$, the theoretical values of $g_{_{\mathrm{VR}}}^{(1)}=51$ and $g_{_{\mathrm{VR}}}^{(2)}=113$, while the numerically predicted values are $47$ and $112$, respectively. $Q$ becomes a minimum at $g=g_{\mathrm{c}}$ at which $\omega_{{\mathrm{r}}}^2$ is minimum and the number of equilibrium points changes from three to one. Essentially, at this value of $g(=g_{\mathrm{c}})$ the effective potential of the slow variable $X$ changes from a double-well form to a single-well.

From the theoretical expression of $Q$, it is possible to determine the values of $g$ at which resonance occurs. From Eq.~(\ref{eq16}) we infer that $Q$ becomes maximum when $\mathrm{d}S/\mathrm{d}g=0$. This condition leads to the following results.

\noindent (i) Resonance occurs at
\begin{subequations}
\label{eq19}
\begin{eqnarray}
    g_{_{\mathrm{VR}}}^{(1)}
      & = & \left[ \frac{\mu^2}{3\beta} \left( 2\vert \omega_0^2 \vert 
                - \omega^2  + \gamma (a-3) \right)
               \right]^{1/2} < g_{\mathrm{c}} , \\
    g_{_{\mathrm{VR}}}^{(2)}
      & = & \left[ \frac{2\mu^2}{3\beta} \left( \vert \omega_0^2 \vert
                 + \omega^2 - \gamma a \right)
                 \right]^{1/2} > g_{\mathrm{c}}
\end{eqnarray}
\end{subequations}
provided
\begin{equation}
\label{eq20}
   \gamma_{\mathrm{c1}}  
     = \frac{\omega^2}{a -1} < \gamma < \gamma_{\mathrm{c2}} 
     = \frac{2 \vert \omega_0^2 \vert -\omega^2}{3-a} . 
\end{equation}
This condition assures that $g_{\mathrm{c}}>0$, $g_{_{\mathrm{VR}}}^{(1)}$,  $g_{_{\mathrm{VR}}}^{(2)}>0$ and $g_{_{\mathrm{VR}}}^{(1)} < g_{\mathrm{c}} < g_{_{\mathrm{VR}}}^{(2)}$. The first resonance takes place at a value of $g(=g_{_{\mathrm{VR}}}^{(1)})<g_{\mathrm{c}}$ while the second resonance is at an another value $g_{_{\mathrm{VR}}}^{(2)}>g_{\mathrm{c}}$. At $g=g_{\mathrm{c}}$ the response amplitude attains a local minimum. The resonances $g_{_{\mathrm{VR}}}^{(1)}$ and $g_{_{\mathrm{VR}}}^{(2)}$ given by Eqs.~(\ref{eq19}) are due to the matching of $\omega_{{\mathrm{r}}}^2$ with $\omega^2- \gamma a$.

\begin{figure}[t]
\begin{center}
\epsfig{figure=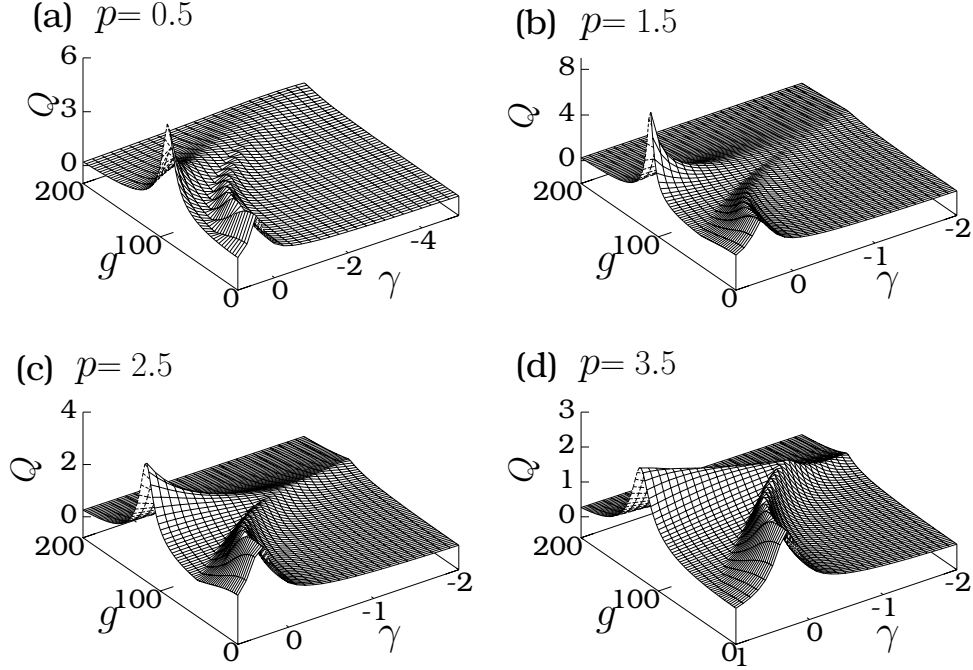, width=0.90\columnwidth}
\end{center}
\caption{Variation of $Q$ as a function of $\gamma$ and $g$ for four fixed values of $p$.  The values of the other parameters are as in Fig.~{\ref{f1}}.}
 \label{f2}
\end{figure}

\noindent(ii) For $\gamma>\gamma_{\mathrm{c2}}$ there is only one resonance at $g=g_{_{\mathrm{VR}}}^{(1)}=g_{\mathrm{c}}$. Here, the maximum value of $Q$ at $g=g_{\mathrm{c}}$ is due to the local minimization of the function $S$ in Eq.~(\ref{eq16}) for which $\omega_{{\mathrm{r}}}^2-\omega^2+\gamma a \ne 0$.\\
(iii) For $\gamma<\gamma_{\mathrm{c1}}$ only one resonance is possible  and it takes place at $g_{_{\mathrm{VR}}}^{(2)}$. Fig.~\ref{f2} presents the influence of the parameters $\gamma$ and $g$ on $Q$ for four fixed values of the parameter $p$ appearing in the gamma distributed delays ( Eqs.~(\ref{eq7}) and (\ref{eq5}).
For $p=0.5$, the conditions (\ref{eq20}) are satisfied for $\gamma \in [-4.49843, 0.42857]$. That is, two resonances can occur when $g$ is varied from a small value for values of $\gamma$ in the above interval. Fig.~\ref{f2}a confirms this. For $p=3.5$ two resonance peaks can take place for $\gamma \in [-0.78,0.31]$. In Fig. {\ref{f2}}, for all values of $p$ the values of $Q$ at resonance decrease with a decrease in the value of $\gamma$ from the value $1$. As $\gamma$ increases from $\gamma_{\mathrm{c1}}$, both $g_{_{\mathrm{VR}}}^{(1)}$ and $g_{_{\mathrm{VR}}}^{(2)}$ move away from $g_{\mathrm{c}}$. 

\begin{figure}[t]
\begin{center}
\epsfig{figure=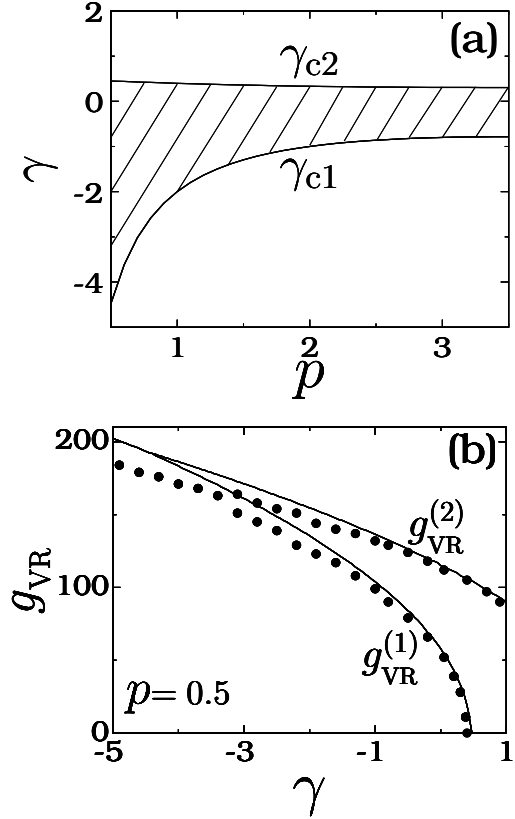, width=0.45\columnwidth}
\end{center}
\caption{(a) $\gamma_{\mathrm{c1}}$ and $\gamma_{\mathrm{c2}}$ versus $p$ for the double-well potential Duffing oscillator. Two resonances take place in the stripped region while only one resonance occurs in the remaining regions. (b) $g_{_{\mathrm{VR}}}$ versus $\gamma$ for  $p=0.5$. The continuous curve and the solid circles are theoretically and numerically computed values of $g_{_{\mathrm{VR}}}$, respectively.}
 \label{f3}
\end{figure}

In Fig.~{\ref{f3}a} the variation of $\gamma_{\mathrm{c1}}$  and $\gamma_{\mathrm{c2}}$ with $p$ is plotted. In the stripped regions two resonances occur. 
For the values of $\gamma$ and $p$ above the curve $\gamma_{\mathrm{c2}}$, there is only one resonance at  $g=g_{_{\mathrm{VR}}}^{(2)}>g_{\mathrm{c}}$ given by Eq.~(\ref{eq19}b) while for below the curve $\gamma_{\mathrm{c1}}$ a single resonance occurs at  $g=g_{\mathrm{c}}$. The Figs.~{\ref{f2}} and {\ref{f3}a} clearly demonstrate the strong influence of the parameters of the delay feedback term. In Fig.~{\ref{f3}b} theoretical and numerically computed  $g_{_{\mathrm{VR}}}$ are shown for a range of values of $\gamma$ with $p=0.5$. The theoretical $g_{_{\mathrm{VR}}}$ matches closely with the numerical $g_{_{\mathrm{VR}}}$ for $\vert \gamma \vert <1$. For  $\vert \gamma \vert >1$ the deviation between the theoretical $g_{_{\mathrm{VR}}}$ and the numerical $g_{_{\mathrm{VR}}}$ increases with an increase in $\vert \gamma \vert $.

\begin{figure}[t]
\begin{center}
\epsfig{figure=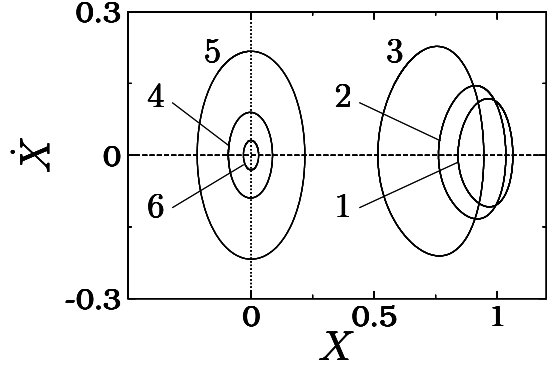, width=0.4\columnwidth}
\end{center}
\caption{Phase portrait of the slow motion $X(t)$ of the system {(\ref{eq6})} for different values of $g$ with $d=0.5$, $\omega_0^2=-1$, $\beta=1$, $f=0.1$, $\omega=1$, $\Omega=10$, $p=0.5$, and $\gamma=0.1$. The values of $g$ for the orbits $1-6$ are: $g=1(1)$, $25(2)$, $47(3)$, $78(4)$, $112(5)$, $185(6)$, respectively.}
 \label{f4}
\end{figure}

Figure~{\ref{f4}} depicts the change in the slow motion $X(t)$ as a function of $g$, computed by solving Eq.~(\ref{eq12}) for the values of the parameters used in Fig.~\ref{f1} with $\gamma=0.1$. 
For $p=0.5$ and $\gamma=0.1$, the values of $g_{_{\mathrm{VR}}}^{(1)}$, $g_{\mathrm{c}}$ and $g_{_{\mathrm{VR}}}^{(2)}$ are $47$, $78$, and $112$, respectively. For $g<g_{\mathrm{c}}$ there are two orbits with one centered at $X_{+}^*$ and another centered at $X_{-}^*$ (not shown in Fig.~\ref{f4}). The $Q$ of both orbits are the same. In Fig.~\ref{f4}, the orbits numbered as $1$, $2$ and $3$ are for $g=1$, $25$ and $47 (=g_{_{\mathrm{VR}}}^{(1)})$. These are the orbits with $X_{+}^*$ as the center. The orbits with $X_{-}^*$ as the center are not shown in Fig.~\ref{f4}. We can clearly see that $Q$ of the orbit-$3$ is the maximum. Further, $X_{+}^*$ (the center of the orbit) moves towards origin with increase in the value of $g$. At $g=g_{\mathrm{c}}=78$, the center of the orbit (orbit-$4$) becomes $X_{0}^*=0$. As $g$ increases further, the center of the orbit remains the same, however, $Q$ increases, reaches a maximum at  $g=g_{_{\mathrm{VR}}}^{(2)}=112$ (orbit-$5$) and then decreases (orbit-$6$ with $g=185$).

\begin{figure}[t]
\begin{center}
\epsfig{figure=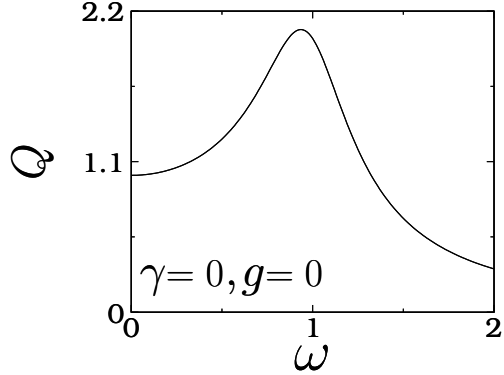, width=0.4\columnwidth}
\end{center}
\caption{Variation of $Q$ with $\omega$ in the absence of time-delayed feedback and high-frequency force for the single-well Duffing oscillator with $d=0.5$, $\omega_0^2=1$, $\beta=1$ and $f=0.1$.}
 \label{f5}
\end{figure}

Next, consider the system (\ref{eq6}) with a single-well potential ($\omega_0^2, \beta>0$). In absence of delay and an external periodic force, the system (\ref{eq12}) has only one real equilibrium point $X_{0}^*=0$. The theoretical expression for $g_{_{\mathrm{VR}}}$ is given by 
\begin{equation}
\label{eq21}
   g_{_\mathrm{VR}}
     = \left[ \frac{2 \mu^2}{3\beta}
         \left( \omega^2 - \omega_0^2 - \gamma a \right)
          \right]^{1/2}. 
\end{equation}
At most one resonance is possible for $\omega^2 - \omega_0^2- \gamma a > 0$ and no resonance for $\omega^2 - \omega_0^2- \gamma a \le 0$. Fig.~\ref{f5} shows $Q$ versus $\omega$ for $\gamma=0$, $g=0$ with $d=0.5$, $\omega_0^2=1$, $\beta=1$ and $f=0.1$.
This figure shows the typical nonlinear resonance due to the applied single frequency periodic force. For a range of values of $\omega$, the response amplitude $Q$ is much lower than its value at resonance. For example, at $\omega=2$ the value of $Q=0.31623$ is lower than the value $Q=2.06527$ corresponding to $\omega=\omega_{\mathrm{max}}=0.94$. The value of $Q$ at $\omega=2$ can be enhanced and resonance can also be realized by including the time-delayed feedback and the high-frequency force.

\begin{figure}[t]
\begin{center}
\epsfig{figure=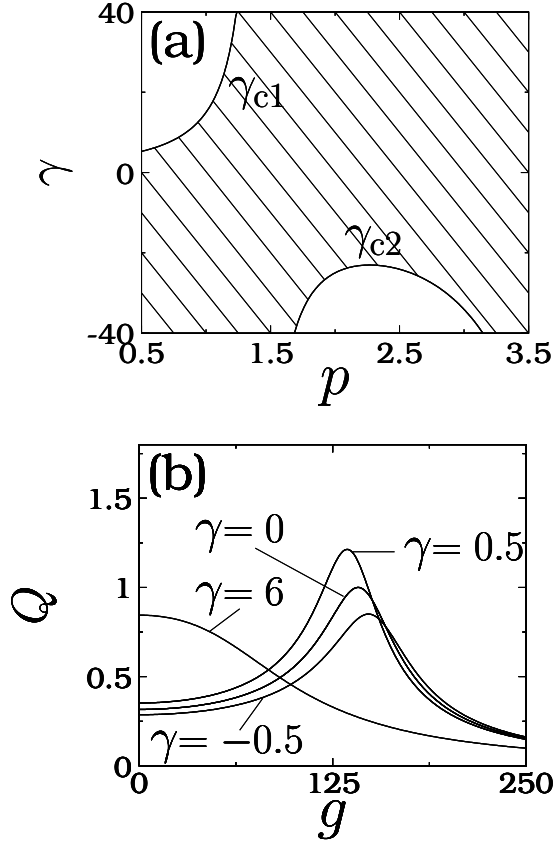, width=0.45\columnwidth}
\end{center}
\caption{(a) Resonance region (stripped region) in the $\gamma$ versus $p$ parameter space for the single-well Duffing oscillator with $d=0.5$, $\omega_0^2=1$,  $\beta=1$, $f=0.1$, $\Omega=10$ and $\omega=2$. (b) $Q$ versus $g$ for four fixed values of $\gamma$ with $p=0.5$.}
 \label{f6}
\end{figure}

From Eq.~(\ref{eq21}) the condition for resonance for $\omega_0^2=1$ and $\omega=2$ is $3-\gamma a>0$. If $a>0$ $(a<0)$, then a resonance occurs only for $\gamma<\gamma_{\mathrm{c1}}=3/a~(\gamma>\gamma_{\mathrm{c2}}=-3/ \vert a \vert)$. Fig.~\ref{f6}a depicts the plot of $\gamma_{\mathrm{c1}}$  and $\gamma_{\mathrm{c2}}$. In the stripped region a resonance can takes place.  
For $p=0.5$, the value of $a$ is $0.56886$ and hence a resonance can occur, when $g$ is varied for $\gamma<\gamma_{\mathrm{c1}}=5.27371$. In Fig.~\ref{f6}b for $\gamma=0$, and $\pm 0.5$ a resonance occurs. For $\gamma=6>\gamma_{\mathrm{c1}}$ when $g$ is increased from zero, the response amplitude monotonically decreases with an increase in $g$ as shown in Fig.~{\ref{f6}b} and there is no resonance. In the single-well case also a resonance occurs when $\omega_{{\mathrm{r}}}^2 = \omega^2 - \gamma a $ (Eq.~(\ref{eq16})), where $\omega_{{\mathrm{r}}}^2=C_1 = \omega_0^2 + {3\beta g^2}/(2\mu^2)$. Unlike the double-well system the center of the slow motion ($X(t)$) and the actual motion ($x(t)$) always takes place about the origin.

\section{Integrative Time-Delayed Duffing Oscillator}
In this section we present the results of the investigation of the vibrational resonance in the Duffing oscillator system with an integrative time-delayed feedback term given by Eq.~(\ref{eq4}).
\subsection{Theoretical Response Amplitude}
We follow the theoretical procedure applied for the system with a gamma distributed time-delayed feedback with $x(t)=X(t)+\psi(t,\Omega t)$. We obtain the following results:
\begin{subequations}
\label{eq22}
\begin{eqnarray}
    Y(t) 
      & = &  A_{\mathrm{L}} \cos (\omega t+\Phi), \\
    A_{\mathrm{L}} 
      & = &  \frac{f}{\sqrt{S} }, \quad  Q = \frac{1}{\sqrt{S}} , \\  
    S 
      & = &  \left[ \omega_{\mathrm{r}}^2 
               - \left( \omega^2 - \frac{\gamma}{\omega \alpha}
               \sin \omega \alpha \right) \right]^2
               + \left[ d \omega - \frac{\gamma}{\omega \alpha} 
               \left( 1 - \cos \omega \alpha \right) \right]^2 ,\\     
    \omega_{{\mathrm{r}}}^2 
      & = &   C_1 + {3\beta X^{*2}}, \quad 
              C_1 = \omega_0^2 + \frac{3 \beta g^2}{2\mu^2} , \\ 
    X^* 
      & = &   0, \; \pm \sqrt{-{(C_1+\gamma)}/{\beta}} \; , \\  
    \mu^2 
      & = & \left[ \omega_0^2 
              - \Omega^2 + \frac{\gamma}{\Omega \alpha} 
               \sin \Omega \alpha \right]^2
              + \left[ d \Omega - \frac{\gamma}{\Omega \alpha} 
              ( 1 - \cos \Omega \alpha ) \right]^2 .  
\end{eqnarray}
\end{subequations}
Notice the difference between the $S$'s given by Eqs.~(\ref{eq16}) and (\ref{eq22}c). Since the solution $Y(t)$ is assumed to be periodic with period $2 \pi/\omega$, we choose $0<\alpha<2 \pi / \omega$. 

\subsection{Double-Well Potential System}
For $\omega_0^2<0$, $\beta>0$ and $\gamma < \vert \omega_0^2 \vert$ the values of $g$ at which the number of equilibrium points changes from three to one and the resonance occurs are given by
\begin{eqnarray}
\label{eq23}
    g_{\mathrm{c}} 
      & = &  \left[ \frac{2\mu^2}{3\beta}            
              \left( \vert \omega_0^2 \vert - \gamma  \right)
               \right]^{1/2}
\end{eqnarray}
and 
\begin{eqnarray}
\label{eq24}
    g_{_{\mathrm{VR}}}^{(1)}
      & = &  \left[ \frac{\mu^2}{3\beta} \left( 2\vert \omega_0^2 \vert 
             - 3 \gamma - \omega^2 
              + \frac{\gamma}{\omega \alpha} \sin \omega \alpha \right)
               \right]^{1/2} < g_{\mathrm{c}} , \\
\label{eq25}
    g_{_{\mathrm{VR}}}^{(2)}
      &  = &  \left[ \frac{2\mu^2}{3\beta} \left( \vert \omega_0^2 \vert
               + \omega^2 - \frac{\gamma}{\omega \alpha} 
                \sin \omega \alpha \right) \right]^{1/2} 
                  > g_{\mathrm{c}} .
\end{eqnarray}
For two resonances to occur the condition on $\gamma$ is $\gamma_{\mathrm{c1}} <  \gamma < \gamma_{\mathrm{c2}}$ where
\begin{equation}
\label{eq26}
   \gamma_{\mathrm{c1}}
     =  - \frac{\omega^2}{1-\frac{1}{\omega \alpha} \sin \omega \alpha},
             \quad \gamma_{\mathrm{c2}}
     =   \frac{2 \vert \omega_0^2 \vert - \omega^2}{3 
             - \frac{1}{\omega \alpha} \sin \omega \alpha} .  
\end{equation}
When  $\gamma > \gamma_{\mathrm{c2}}$, only one resonance is possible with the corresponding  $g_{_{\mathrm{VR}}}$ given by $g_{_{\mathrm{VR}}}^{(2)}$. For $\gamma < \gamma_{\mathrm{c1}}$ there will be only one resonance at $g = g_{\mathrm{c}}$.

\begin{figure}[t]
\begin{center}
\epsfig{figure=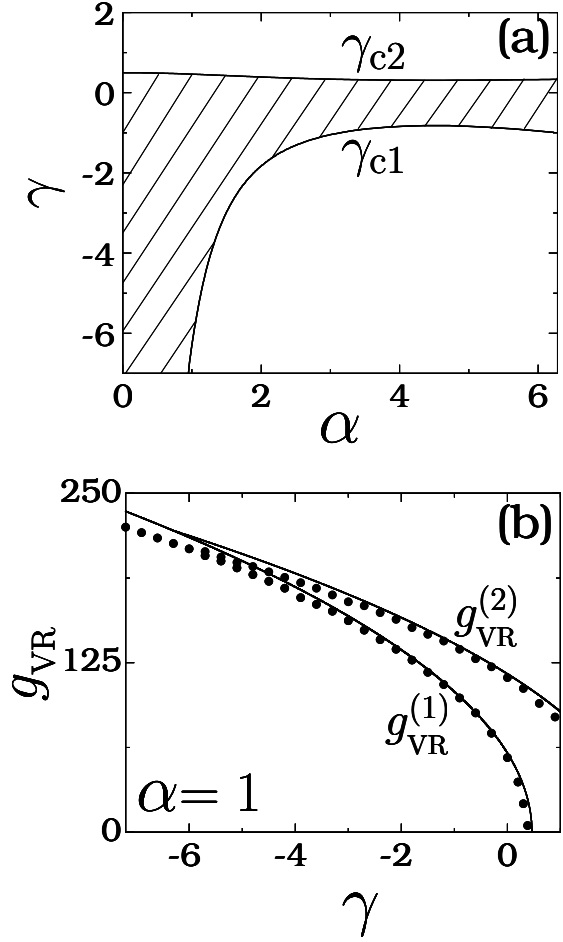, width=0.47\columnwidth}
\end{center}
\caption{(a) Dependence of $\gamma_{\mathrm{c1}}$ and $\gamma_{\mathrm{c2}}$ with the time-delay $\alpha$ for the system {(\ref{eq6})} with an integrative time-delayed feedback term. Here $d=0.5$, $\omega_0^2=-1$, $\beta=1$, $f=0.1$, $\omega=1$ and $\Omega=10$. Two resonances occur in the stripped region. Below and above the stripped region only one resonance is possible. (b) Theoretically predicted (continuous curve) and numerically computed (solid circles) values of  $g_{_{\mathrm{VR}}}^{(1)}$ and  $g_{_{\mathrm{VR}}}^{(2)}$ for $\alpha=1$ as a function of the parameter $\gamma$.}
 \label{f7}
\end{figure}

Figure~\ref{f7}a depicts the threshold curves $\gamma_{\mathrm{c1}}$ and $\gamma_{\mathrm{c2}}$ for $\omega_0^2=-1$, $\beta =1$, $d=0.5$,  $f=0.1$, $\omega=1$ and $\Omega=10$. In the stripped region, two resonances occur while only one resonance occurs in the remaining region. In Fig.~\ref{f7}b, both theoretically predicted and numerically computed $g_{_{\mathrm{VR}}}^{(1)}$ and $g_{_{\mathrm{VR}}}^{(2)}$ are plotted as a function of $\gamma$ for $\alpha=1$. Here again, the theoretical result closely matches with the numerical simulation.

For $\alpha=1$, we find $\gamma_{\mathrm{c1}}=-6.308$ and $\gamma_{\mathrm{c2}}=0.46394$. Two resonances are possible for $-6.308< \gamma < 0.46394$ and only one resonance can occur outside this interval of $\gamma$, when $g$ is varied. When $\gamma=0.2$,  $\omega_{{\mathrm{r}}}^2$ is equal to $\omega^2-(\gamma \sin \omega \alpha)/\omega \alpha$ at two values of $g$, namely at $g=43.5$ and $g=113.55$. At these two values of $g$, $Q$ becomes maximum. $\omega_{{\mathrm{r}}}^2$ matches with $\omega^2-(\gamma \sin \omega \alpha)/\omega \alpha$ at only one value of $g$ for $\gamma=0.5$ and hence only one resonance. Here, the resonance occurs at $g=103.5 > g_{\mathrm{c}}=59$.

\begin{figure}[t]
\begin{center}
\epsfig{figure=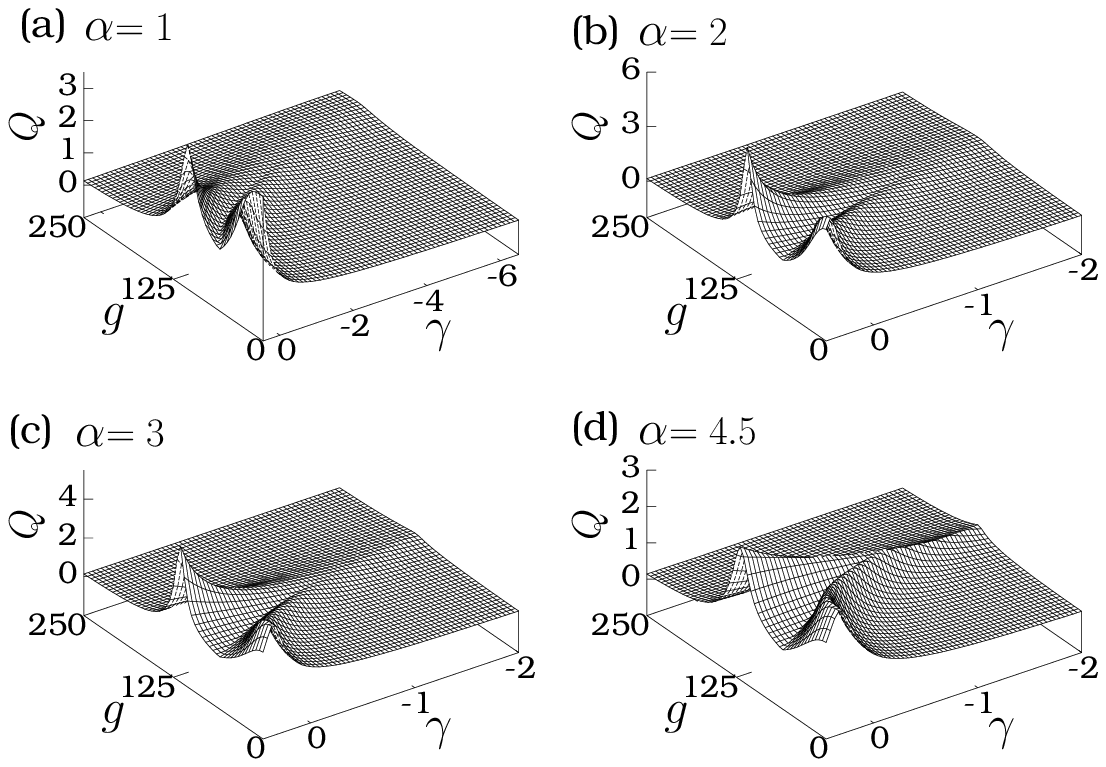, width=0.999\columnwidth}
\end{center}
\caption{$Q$ versus $g$ and $\gamma$ for four fixed values of $\alpha$ for the double-well system {(\ref{eq6})} with an integrative time-delayed feedback term.}
 \label{f8}
\end{figure}

The influence of the parameters $g$ and $\gamma$ on $Q$ is depicted in Fig.~\ref{f8} for four fixed values of time-delay $\alpha$. In this figure, we clearly notice that for $\gamma$ values below a critical value $(\gamma_{\mathrm{c1}})$ only one resonance takes place. Comparing the Fig.~\ref{f2} of the gamma distributive time-delayed feedback case and Fig.~\ref{f8} of the integrative time-delayed feedback case, the effect of $\alpha$ is found to be similar to that of $p$.

%
\subsection{Single-well Potential System}  
\begin{figure}[t]
\begin{center}
\epsfig{figure=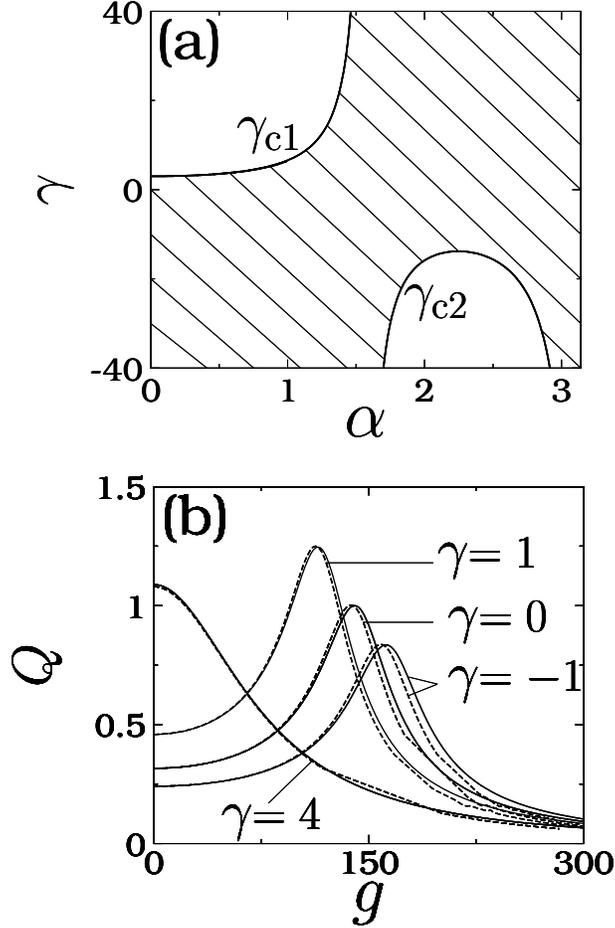, width=0.5\columnwidth}
\end{center}
\caption{(a) Single resonance region (stripped region) and no resonance region (blank region) for the single-well potential of the system (\ref{eq6}) with an integrative time-delayed feedback term. $\gamma_{\mathrm{c1}}$ and $\gamma_{\mathrm{c2}}$ are given by Eqs.~(\ref{eq25}). Here $\omega_0^2=1$ and $\omega=2$. (b) $Q$ versus $g$ for four fixed values of $\gamma$ with $\alpha=0.2$. The continuous and dashed lines are theoretically and numerically computed values of $Q$, respectively.}
 \label{f9}
\end{figure}

For the single-well potential case $(\omega_0^2,\beta > 0)$, the slow motion takes place about the equilibrium point $X^*=0$ and the theoretically predicted $g_{_{\mathrm{VR}}}$ is 
\begin{equation}
\label{eq27}
    g_{_{\mathrm{VR}}}
     =  \left[ \frac{2\mu^2}{3\beta} \left( - \omega_0^2 
            +\omega^2 - \frac {\gamma}{\omega \alpha}  \sin \omega \alpha \right)
            \right]^{1/2}.
\end{equation}
From (\ref{eq27}), we find the condition for resonance as 
\begin{equation}
\label{eq28}  
    - \omega_0^2 + \omega^2 
       - \frac{\gamma}{\omega \alpha} \sin \omega \alpha > 0 .            
\end{equation}
We fix $d=0.5$, $\omega_0^2=1$, $\beta=1$, $f=0.1$, $\omega=2$ and $\Omega=10$. In this case, when $g$ is varied $Q$ can exhibit a resonance only if 
\begin{subequations}
 \label{eq29}
\begin{eqnarray}
   \gamma < \gamma_{\mathrm{c1}}
      = \frac{6 \alpha}{\sin 2 \alpha}, \quad
           \alpha \in \left[ 0, \pi/2 \right]  
\end{eqnarray} 
or   
\begin{eqnarray}
    \gamma > \gamma_{\mathrm{c2}}
      =  \frac{-6 \alpha}{\vert \sin 2 \alpha \vert},
          \quad \alpha \in \left[ \pi/2, \pi \right ] . 
\end{eqnarray}
\end{subequations} 
where the range of $\alpha$ is restricted to $\left[ 0, 2 \pi/\omega \right] = \left[0, \pi \right]$. The threshold curves $\gamma_{\mathrm{c1}}$ and $\gamma_{\mathrm{c2}}$ are plotted in Fig.~\ref{f9}a. In the stripped region, only one resonance can occur. In the remaining set of values of $\gamma$ and $\alpha$, the integrative time-delay suppresses the existing resonance for $\gamma=0$. When $\alpha=0.2<\pi$, the theoretically predicted $\gamma_{\mathrm{c1}}=3.08$, while the numerically computed $\gamma_{\mathrm{c1}}=3.04$. In Fig.~\ref{f9}b both theoretically and numerically calculated $Q$ are shown for four values of $\gamma$. For $\gamma=\pm 1 <\gamma_{\mathrm{c1}}$, $Q$ displays a single resonance. For $\gamma= 4 > \gamma_{\mathrm{c1}}$, the response amplitude monotonically decreases with $g$ and there is no resonance. 

%
\section{Conclusions}
In this paper, we have analyzed the influence of distributed time-delays and integrative time-delays on vibrational resonance for the Duffing oscillator, by using a theoretical approach. Using the approximate analytical expression for the response amplitude, we have been able to determine the parameter regions (describing the time-delayed feedback) where either two resonances, one single resonance or no resonance occur. An enhanced response is realized for a range of values of the control parameters of the feedback. With and without time-delayed feedback at most two resonances and at least one resonance occur in the case of the double-well Duffing oscillator. In the single-well case, at most one resonance is possible. Suppression of this single resonance by time-delayed feedback happens for a range of values of the parameters. Thus, the time-delayed feedback can be used to control the number of resonances and the value of the response amplitude at the resonance.

 \vskip 5pt
\noindent{\bf{Acknowledgment}}
 \vskip 5pt
MAFS acknowledges financial support from the Spanish Ministry of Economy and Competitivity under Project No. FIS2013-40653-P.

\appendix
\numberwithin{equation}{section}
\section{ }

{\centerline{\bf{PROOF OF EQUATION (\ref{eq10}) }}
We present the details of the evaluation of the term $F(\tau, \psi(\Omega t-\Omega \tau))$ given by Eq.~(\ref{eq7}) for $\psi=A_{\mathrm{H}} \cos(\Omega t + \phi)$. Equation~(\ref{eq7}) with the $\psi$ given above takes the form
\begin{equation}
 \label{eqA1}
    F  = \frac{\gamma A_{\mathrm{H}}}{\Gamma{(p)}}  
          \int_0^{\infty} {\tau^{p-1}} {\mathrm{e}}^{- \tau } 
           \cos(\Omega t + \phi - \Omega \tau)\, {\mathrm{d}} \tau .   
\end{equation}  
Using $\cos \theta = {\mathrm {Re}} \left ({\mathrm{e}}^{{\mathrm{i}} \theta}\right)$ the above equation is rewritten as 
\begin{equation}
\label{eqA2}
     F = {\mathrm{Re}} \left[\frac {\gamma A_{\mathrm{H}}}{{\Gamma{(p)}}} 
           {\mathrm{e}}^{{\mathrm{i}} (\Omega t +\phi)} 
           \int_0^{\infty} {\tau^{p-1}} \, {\mathrm{e}}^{- b \tau }
            \, {\mathrm{d}} \tau \right] ,   
\end{equation} 
where $b=1+{\mathrm{i}} \Omega$. As the value of the integral in (\ref{eqA2}) is ${\Gamma{(p)}}/b^{p}$, we have 
\begin{eqnarray}
\label{eqA3}
   F & = &  {\mathrm{Re}} \left[ \frac{\gamma A_{\mathrm{H}} 
            {\mathrm{e}}^{{\mathrm{i}} (\Omega t +\phi)}} 
            {{(1+{\mathrm{i}} \Omega)}^p } \right], \\  
  \label{eqA4}
    & = & {\mathrm{Re}} \left[ {\gamma A_{\mathrm{H}} 
            {\mathrm{e}}^{{\mathrm{i}} (\Omega t +\phi)}} \,
           {\mathrm{i}}^{-p} \, \Omega^{-p} \, 
           {\left( 1-\frac {{\mathrm{i}} }{\Omega} \right) }^{-p} 
            \right].  
\end{eqnarray}
For $\Omega \gg 1$, $\left( 1 - \frac{ {\mathrm{i}} }{ \Omega } \right)^{-p} \approx  \left( {\mathrm{e}}^{ -{\mathrm{i}} / \Omega } \right)^{-p} = {\mathrm{e}}^{ {\mathrm{i}} p / \Omega}$. Then
\begin{eqnarray}
 \label{eqA5}
    F 
      & = &  {\mathrm{Re}} \left[ \gamma  A_{\mathrm{H}}  \Omega^{-p} 
             {\mathrm{e}}^{ {\mathrm{i}} ( \Omega t 
             + \phi + p/\Omega ) }  
             {\mathrm{e}}^{-{\mathrm{i}} p \pi/2} \right]
             \nonumber \\
      & = & \gamma  A_{\mathrm{H}}  \Omega^{-p} 
              \cos(\Omega t + \phi + \theta ) , 
\end{eqnarray} 
where ${\displaystyle{\theta = \frac {p}{\Omega} - \frac {p \pi}{2} }}$.

\end{document}